\documentclass[aps,pra,preprint,groupedaddress,floatfix]{revtex4}
\usepackage{graphics}

\begin{document}
 
\preprint{LA-UR-07-4046}
 
\title{Multiple Extremal Eigenpairs by the Power Method}
 
 
\author{J. E. Gubernatis}
\affiliation{Theoretical Division, Los Alamos National Laboratory, Los Alamos, NM 87545 U.S.A.}
\author{T. E. Booth}
\affiliation{ Applied Physics Division, Los Alamos National Laboratory, Los Alamos, NM 87545 U.S.A.}

 
\date{\today}
 
\begin{abstract}

We report the production and benchmarking of several refinements of the power method that enable the computation of multiple extremal eigenpairs of very large matrices. In these refinements we used an observation by Booth that has made possible the calculation of up to the 10$^{th}$ eigenpair for simple test problems simulating the transport of neutrons in the steady state of a nuclear reactor. Here, we summarize our techniques and efforts to-date on determining mainly just the two largest or two smallest eigenpairs. To illustrate the effectiveness of the techniques, we determined the two extremal eigenpairs of a cyclic matrix, the transfer matrix of the two-dimensional Ising model, and the Hamiltonian matrix of the one-dimensional Hubbard model.

\end{abstract}
 
 
\maketitle
 
 
\section{Introduction}

Computing eigenpairs of large matrices is a ubiquitous problem in computational physics. In this paper, we present several refinements of the basic power method that enable the efficient and accurate computation of multiple extremal eigenvalues of very large matrices.  Ultimately, our objective is producing Monte Carlo versions of such methods for matrices whose orders are so large that even the eigenvectors cannot be stored in computer memory. For such problems, the computation of a basic vector quantity as the inner product is generally either very inefficient or impractical. It can be impractical, for instance, because the nature of Monte Carlo sampling means most components of these vectors are unknown. Here, we focus on the basic algorithms developed to date, noting they work well when used deterministically. Novel will be the illustration of how the power method can be expanded to compute  several extremal eigenpairs simultaneously rather than just one at a time. While various versions of the power method often compute very well the dominant eigenvalue $\lambda _1 $, the one with largest absolute value, computing subdominant eigenvalues $ \lambda _2 ,\lambda _{3,} \ldots $ has often proven much more difficult and is much less frequently attempted. 

The algorithms to be presented use some recent insights of Booth \cite{booth1,booth2} that were developed for Monte Carlo simulations of steady state neutron transport in nuclear reactors. Initially, he proposed a novel modification of the power method that has produced up to 10 eigenpairs for simple test problems. Here, we present refinements of these insights and focus on determining just $\lambda _1 $ and $ \lambda _2 $, plus their eigenvectors. The convergence of the power method is well known to slow as the ratio $\lambda _2 /\lambda _1$, sometimes called the dominance ratio, approaches unity.  As such, this ratio is an indicator of solution difficulty and acceptability. As we will illustrate, an advantage of computing two dominant eigenpairs simultaneously is often improved convergence to the first one. An advantage of the present techniques is the ease in getting both eigenfunctions along with their eigenvalues.

It is important to note that various areas of science and engineering seek multiple eigenpairs for reasons other than algorithmic gains. In nuclear engineering, a dominance ratio distinct from unity is an acceptance qualifier for various nuclear criticality safety assessments and nuclear reactor designs \cite{spanier}. In statistical physics, a dominance ratio nearing unity, on the other hand,  is often a condition sought. Near a continuous phase transition, $ \lambda _2  \to \lambda _1 $,  and $ \lambda_2/\lambda_1 $ controls the microscopic spatial correlations among {\it physical\/} degrees of freedom \cite{thompson}. Today, an important topic in quantum statistical mechanics is quantum critical phenomena, phase transitions driven by zero-point motion at zero temperature \cite{sachdev}. Here, it is the two smallest eigenvalues of the Hamiltonian matrix describing the physical system that are of interest. The quantum critical phenomenon construct, while supplemented by a few exact solutions to some very simple problems, is largely phenomenological in part because of the inability to compute $\lambda_2$ for models of direct physical relevance.

In the next section, Section~II, we summarize some basic notions about the power method and our refined procedures. For simplicity, we will assume the two largest extremal eigenpairs are sought. Also we restrict attention to systems  with real eigenvectors and eigenvalues, but our methods can be applied to complex systems as well. In Section~III, we apply these techniques to determination of a few eigenpairs of three problems. The first is the cyclic matrix that results from the discretization of the gradient operator on a circle, the second is the transfer matrix of the two-dimensional Ising moel, and the third is the Hamiltonian matrix of the one-dimensional Hubbard model. For the first and third problems, we determine the smallest two eigenpairs (ground-state and first excited state) instead of the largest ones to illustrate the flexibility of the techniques. The second and third problems counterpose in their computational challenges in a number of ways: The transfer matrix for the Ising model is non-symmetric, positive, and dense. Its eigenvalues are known analytically and all its matrix elements follow a single simple analytic expression. The Hamiltonian matrix for the Hubbard model is symmetric, indefinite, and sparse. Its two smallest eigenvalues are not known analytically, and its matrix elements, while easy to compute, lack a simple expression. In Section~IV, the final section, we will discuss extensions of the techniques to broader classes of problems, including those involving continuous operators.

\section{Methodology}

We first summarize the power method, and then we discuss ways to refine it so convergence is to the two largest extremal eigenpairs simultaneously. We conclude this section with two refinements of the power method: one is necessary for the reduction of round-off error and the other improves the convergence rate to the dominant eigenpair while simultaneously calculating the second extremal eigenpair.

\subsection{Power Method Basics}

For some real-valued $N\times N$ matrix $A$, not necessarily symmetric, we will be concerned with the $N$ eigenpairs $(\lambda_i,\psi_i)$ satisfying
\begin{equation}
A\psi_i  = \lambda _i \psi_i
\label{eq:eigenvalue}
\end{equation}
In the simplest application of the power method \cite{wilkinson}, an iteration is started with
some  $\psi$, normalized in some convenient, but otherwise relatively
arbitrary, manner and consists of iterating two steps
\begin{equation}
\begin{array}{c}
 \phi = A\psi  \\
 \psi = \phi / \|\phi\|\\
 \end{array}
\label{eq:power_method}
\end{equation}
If we write
\[
\psi= \sum\limits_{i = 1}^N {\alpha _i \psi_i }
\]
then after $n$ iterations
\begin{equation}
A^n\psi  = \lambda _1^n \left[ {\alpha _1 \psi_1  + \sum\limits_{i = 2}^N {\alpha _i \left( {\frac{{\lambda _i }}
{{\lambda _1 }}} \right)^n \psi_i } } \right]
\label{eq:pm}
\end{equation}
If $\left| {\lambda _1 } \right| > \left| {\lambda _2 } \right| \ge \left|{\lambda _3 } \right| \ge \cdots  \ge \left| {\lambda _N } \right|$, then for $\alpha_1 \neq 0$,
\[
\begin{array}{c}
 \psi  \to \psi_1 /\|\psi_1\| \\
 \|\phi\| \to \lambda _1  \\
 \end{array}
 \]
Hence, the dominant eigenpair is simultaneously determined. For the norm of the vector $\phi$ whose components are $\phi_i$, a frequent choice is
\[
\parallel \phi \parallel \equiv \parallel \phi \parallel_\infty = \max_i |\phi_i|
\]
This is the choice adopted here.

Clearly, if $\left| {\lambda _2 /\lambda _1 } \right| \simeq 1$ convergence of the iteration is slow. Often it can be improved by the replacement $A \to A - \sigma I$ which shifts the value of each eigenvalue by a constant amount $\sigma$ but does not change the associated eigenvector. Besides potentially accelerating convergence, the shift also enables the determination of the smallest, instead of the largest, eigenpair. In particular, if $A$ and all the $\lambda_i$ are real, no matter how $\sigma$ is chosen, either $\lambda_1-\sigma$ or $\lambda_N-\sigma$ will be the converged eigenvalue. Most often, $\sigma$ is chosen to  be independent of iteration step. In this case, for convergence to $\lambda_1$, the optimal choice for $\sigma$ is $\frac{1}{2}\left( {\lambda _2  + \lambda _N } \right)$; for convergence to $\lambda_N$, the choice is $\frac{1}{2}\left( {\lambda _1  + \lambda _{N-1} } \right)$ \cite{wilkinson}.
  
If the dominant eigenvalue is degenerate, for example, doubly degenerate with $\lambda_1=\lambda_2$, or degenerate in magnitude, for example, doubly degenerate with $|\lambda_1|=|\lambda_2|$, then the power method, as most iterative methods,  cannot determine a unique eigenvector. As can be seen from Eq. (\ref{eq:pm}), in these situations the iteration converges to 
\[
A^n\psi  = \lambda _1^n \left[ {\alpha _1 \psi_1  + {\rm sign}\left(\frac{\lambda_2}{\lambda_1}\right)\alpha_2\psi_2 + \sum\limits_{i = 3}^N {\alpha _i \left( {\frac{{\lambda _i }}
{{\lambda _1 }}} \right)^n \psi_i } } \right]
\]
The eigenvalue estimators will converge to the correct values of $\lambda_1$ and $\lambda_2$ but the eigenvector estimate corresponding to the dominant eigenvalue  will be some linear combination of $\psi_1$ and $\psi_2$. A similar situation will can occur for convergence to the first subdominant eigenvalue if for example $|\lambda_2|=|\lambda_3|$. In this case $\psi_1$ can be determined but $\psi_2$ cannot.

If a few dominant eigenpairs, say $M$, are desired, one of two approaches are tried. One approach is to use the power method to determine the dominant eigenpair, use deflation to project out this state out of the matrix, and then reuse the power method on the deflated matrix. To determine several eigenpairs simultaneously, the power method can be generalized to 
\[
 \Phi  = A\Psi
\]
where $\Phi$ and $\Psi$ are $M\times N$ matrices whose columns are orthogonalized to each other. This orthogonality needs maintenance throughout the computation or else all $M$ vectors, represented by the columns of the initial $\Psi$, will converge to the same one. This algorithm is called the simultaneous iteration method \cite{golub}.

\subsection{Observation}
  
Booth's refinement of the power method \cite{booth1,booth2} uses the observation that for any eigenpair $(\lambda,\psi)$ and for each non-zero {\it component\/} of the eigenvector, the eigenvalue equation $A\psi  = \lambda\psi$ can be rewritten as
\begin{equation}
\lambda  = \frac{{\sum\limits_\beta  {A_{\alpha \beta } \psi _\beta  } }}{{\psi _\alpha  }}
\label{eq:eigenvalue_estimator}
\end{equation}
and that similar equations can also be written for any number of groupings of components,
\begin{equation}
\lambda  = \frac{{\sum\limits_{\alpha  \in R_1 } {\sum\limits_\beta  {A_{\alpha \beta } \psi _\beta  } } }}{{\sum\limits_{\alpha  \in R_1 } {\psi _\alpha  } }} 
= \frac{{\sum\limits_{\alpha  \in R_2 } {\sum\limits_\beta  {A_{\alpha \beta } \psi _\beta  } } }}{{\sum\limits_{\alpha  \in R_2 } {\psi _\alpha  } }} 
=  \cdots  
= \frac{{\sum\limits_{\alpha  \in R_N } {\sum\limits_\beta  {A_{\alpha \beta } \psi _\beta  } } }}{{\sum\limits_{\alpha  \in R_N } {\psi _\alpha  } }}
\label{eq:groupings}
\end{equation}
where the $R_i$ are rules for different groupings. The groups can overlap.  In addition, any two groupings, say 1 and 2, imply
\begin{equation}
\sum\limits_{\alpha  \in R_2 } {\psi _\alpha  } \sum\limits_{\alpha  \in R_1 } {\sum\limits_\beta  {A_{\alpha \beta } \psi _\beta  } }  = \sum\limits_{\alpha  \in R_1 } {\psi _\alpha  } \sum\limits_{\alpha  \in R2} {\sum\limits_\beta  {A_{\alpha \beta } \psi _\beta  } } 
\label{eq:cross_product}
\end{equation}
The eigenvalue estimator (\ref{eq:eigenvalue_estimator}) is a special case of what is often called a mixed estimator \cite{hammond}
\[
\lambda= \frac{\langle\phi|A|\psi\rangle}{\langle\phi|\psi\rangle}
\]
In the present case, the component $\phi_i$ of the vector $\phi$ is unity if $i\in R$; otherwise, it is zero. From $N$ groupings of the components, Booth constructs $N$ estimators for the $N^{th}$ eigenvalue and forces them to become equal by adjusting certain parameters at each iteration step. Several ways to do this have been devised, and we will now sketch the most recent ones for obtaining two extremal eigenpairs simultaneously.

For almost any starting point $\psi  = \sum\nolimits_i {\alpha _i \psi _i } $, the power method will converge to $ \left( {\lambda _1 ,\psi _1 } \right)$. The same would be true for almost any other two normalized, but not necessarily orthogonal, starting points $\psi'  = \sum\nolimits_i {b _i \psi _i } $ or $\psi'' = \sum\nolimits_i {a _i \psi _i } $. We will in fact choose two such starting points and at each step apply $A$ to them individually, but at each step we will adjust the relationship between them to prevent the collapse of their {\it sum} to the dominant eigenfunction. 

Formally, we start the iteration with $\psi=\psi'+\eta\psi''$ and assume that after a large number of steps just the two dominant eigenpairs remain significant. Then we have
\begin{equation}
A^n \psi  = A^n \sum\limits_{i = 1}^2 {\alpha _i \psi _i }  =
\sum\limits_{i = 1}^2 {\left( {a_i  + \eta b_i } \right)
\lambda_i^n \psi _i}
\label{eq:nth_step}
\end{equation}
To determine $\eta$, we define two groupings of the components of $A^n \psi$, $R_1$ and $R_2$, and let $ \kappa_j $ be the eigenvalue estimate for the $j^{th}$ grouping. Then from Eq.~(\ref{eq:groupings}) we find that
\[
\begin{array}{c}
 \kappa _1  = \frac{\displaystyle {\left( {a_1  + \eta b_1 } \right)\lambda _1^n \sum_{\alpha \in {R_1 }} \psi_{1,\alpha} + \left( {a_2  + \eta b_2 } \right)\lambda _2^n \sum_{\alpha \in {R_1 }} \psi_{2,\alpha} }}{\displaystyle{\left( {a_1  + \eta b_1 } \right)\lambda _1^{n - 1} \sum_{\alpha \in {R_1 }} \psi_{1,\alpha}  + \left( {a_2  + \eta b_2 } \right)\lambda _2^{n - 1} \sum_{\alpha \in {R_1 }} \psi_{2,\alpha} }} \\ 
 \kappa _2  = \frac{\displaystyle{\left( {a_1  + \eta b_1 } \right)\lambda _1^n \sum_{\alpha \in {R_2 }} \psi_{1,\alpha} + \left( {a_2  + \eta b_2 } \right)\lambda _2^n \sum_{\alpha \in {R_2 }} \psi_{2,\alpha} }}{\displaystyle{\left( {a_1  + \eta b_1 } \right)\lambda _1^{n - 1} \sum_{\alpha \in {R_2 }} \psi_{1,\alpha}  + \left( {a_2  + \eta b_2 } \right)\lambda _2^{n - 1} \sum_{\alpha \in {R_2 }} \psi_{2,\alpha} }} \\ 
 \end{array}
\]
If we require $\kappa_1  = \kappa_2 $, a quadratic equation for $\eta$ results. If one solution of this equation is chosen to guide the iteration to $a_1+\eta b_1=0$, then $\kappa_1=\kappa_2=\lambda_2$. If the solution on the other hand is chosen to guide the iteration towards $a_2+\eta b_2=0$, then $\kappa_1=\kappa_2=\lambda_1$. 

In practice, we find the coefficients of this quadratic equation in the following manner: Suppose at the $n^{th}$ step, $\psi'$ and $\psi''$ have iterated to $\hat\psi'$ and $\hat\psi''$, then at the $(n+1)^{th}$ step we require
\begin{equation}
\frac{{\sum\limits_{\alpha  \in R_1 } {\sum\limits_\beta  {A_{\alpha \beta } \hat \psi '_\beta  } }  + \eta \sum\limits_{\alpha  \in R_1 } {\sum\limits_\beta  {A_{\alpha \beta } \hat \psi ''_\beta  } } }}{{\sum\limits_{\alpha  \in R_1 } {\hat \psi '_\alpha  }  + \eta \sum\limits_{\alpha  \in R_1 } {\hat \psi ''_\alpha  } }} = \frac{{\sum\limits_{\alpha  \in R_2 } {\sum\limits_\beta  {A_{\alpha \beta } \hat \psi '_\beta  } }  + \eta \sum\limits_{\alpha  \in R2} {\sum\limits_\beta  {A_{\alpha \beta } \hat \psi ''_\beta  } } }}{{\sum\limits_{\alpha  \in R_2 } {\hat \psi '_\alpha  }  + \eta \sum\limits_{\alpha  \in R_2 } {\hat \psi ''_\alpha  } }}
\label{eq:balance}
\end{equation}
which leads to $q_2\eta^2+q_1\eta+q_0=0$ with
\[
q_2 = \sum\limits_{\alpha  \in R_2 } {\hat \psi ''_\alpha  } \sum\limits_{\alpha  \in R_1 } {\sum\limits_\beta  {A_{\alpha \beta } \hat \psi ''_\beta  } }  - \sum\limits_{\alpha  \in R_1 } {\hat \psi ''_\alpha  } \sum\limits_{\alpha  \in R2} {\sum\limits_\beta  {A_{\alpha \beta } \hat \psi ''_\beta  } } 
\]
\begin{eqnarray}
q_1 &=& \sum\limits_{\alpha  \in R_2 } {\hat \psi ''_\alpha  } \sum\limits_{\alpha  \in R_1 } {\sum\limits_\beta  {A_{\alpha \beta } \hat \psi '_\beta  } }  - \sum\limits_{\alpha  \in R_1 } {\hat \psi ''_\alpha  } \sum\limits_{\alpha  \in R_2 } {\sum\limits_\beta  {A_{\alpha \beta } \hat \psi '_\beta  } } \nonumber \\
    &+&  \sum\limits_{\alpha  \in R_2 } {\hat \psi '_\alpha  } \sum\limits_{\alpha  \in R_1 } {\sum\limits_\beta  {A_{\alpha \beta } \hat \psi ''_\beta  } }  - \sum\limits_{\alpha  \in R_1 } {\hat \psi '_\alpha  } \sum\limits_{\alpha  \in R_2 } {\sum\limits_\beta  {A_{\alpha \beta } \hat \psi ''_\beta  } }
\label{eq:q210} 
\end{eqnarray}

\[
q_0 = \sum\limits_{\alpha  \in R_2 } {\hat \psi '_\alpha  } \sum\limits_{\alpha  \in R_1 } {\sum\limits_\beta  {A_{\alpha \beta } \hat \psi '_\beta  } }  - \sum\limits_{\alpha  \in R_1 } {\hat \psi '_\alpha  } \sum\limits_{\alpha  \in R_2 } {\sum\limits_\beta  {A_{\alpha \beta } \hat \psi '_\beta  } }
\]
The strategy is to apply $A$ repeatedly until two real solutions for $\eta$ exist.  One solution will then guide the iteration to $(\lambda_1,\psi_1)$; the other, to $(\lambda_2,\psi_2)$. Typically, this procedure would be used only if the simultaneous convergence to two pairs is desired or if the convergence to just the second eigenpair is desired. In some cases, however, accelerated convergence to the first pair occurs.

\subsection{First Refinement}

For simplicity, we focus on the determination of the {\it second largest} eigenpair \cite{booth2} and note that one additional improvement is necessary for finite precision computers. As both $\hat\psi'$ and $\hat\psi''$ are converging to the first eigenfunction, only their sum, for proper choices of $\eta$, is converging to the second one. Eventually, when $\eta$ is the root, say $\eta_2$, guiding $\hat\psi''$ toward the second eigenvector $\psi_2$, the determination of $\psi_2$ is limited by the accuracy of the sum of $\hat\psi'$ and $\eta_2\hat\psi''$. To mitigate this situation, we modify the iteration by making the replacements $\hat\psi'\leftarrow \hat\psi'$ and $\hat\psi''\leftarrow \hat\psi''+\eta_2\hat\psi'$ before moving to the $(n+1)^{th}$ step, and then in the $(n+1)^{th}$ step we find the new $\eta$ from the quadratic equation and subtract from it the $\eta_2$ from the $n^{th}$ step. Formally, this is equivalent to rewriting the coefficients of the $\psi_i$ in Eq.~(\ref{eq:nth_step}) as
\begin{equation}
a_i + b_i \eta =(a_i+b_i \eta_2) + b_i (\eta-\eta_2)
\label{eq:further_refinement}
\end{equation}
making the replacements
\[
a_i \leftarrow (a_i+b_i \eta_2)
\]
\[
\eta  \leftarrow \eta-\eta_2
\]
and then in the next iteration solving the quadratic equation for the shifted $\eta$. 

What does this procedure accomplish? We note that near convergence, when only $\psi_1$ and $\psi_2$ remain significant, the current best estimate of $\psi_1$ is contaminated with $\psi_2$ and vice versa. Denoting these estimates by $\psi_1 +\epsilon\psi_2$ and $\psi_2 + \delta\psi_1$ and introducing the adjustable parameter $\eta$, we can write another estimate of $\psi_2(\eta)$ as
\[
 \hat \psi_2(\eta) =  (\psi_2 + \delta \psi_1) + \eta (\psi_1 + \epsilon \psi_2)
\]
and so with the application of $A$ to move to the $(n+1)^{th}$ step, the new estimate of the second eigenfunction becomes
\[
\psi_{2new}=A  \hat \psi_2(\eta) =  (\lambda_2 \psi_2 + \delta \lambda_1\psi_1) + \eta (\lambda_1 \psi_1 + \epsilon \lambda_2 \psi_2)
\]
If at this step $\eta_2$ is the choice that guides to $\psi_2$, then we define $\eta''$ by
\[
\eta=\eta''+\eta_2
\]
so that
\begin{eqnarray*}
\psi_{2new}(\eta') &=&
  (\lambda_2 \psi_2 + \delta \lambda_1 \psi_1) + (\eta''+\eta_2) (\lambda_1 \psi_1 + \epsilon \lambda_2 \psi_2) \\
                  &=&
  \Bigl[
  (\lambda_2 \psi_2 + \delta \lambda_1 \psi_1) 
 +\eta_2 (\lambda_1 \psi_1 + \epsilon \lambda_2 \psi_2) 
 \Bigr]
  + \eta'' (\lambda_1 \psi_1 + \epsilon \lambda_2 \psi_2)
\end{eqnarray*}
We observe that $\lambda_1 \psi_1 + \epsilon \lambda_2 \psi_2$ is this step's power iteration estimate for the first eigenfunction so that the second eigenfunction, the term in brackets, is essentially being corrected by an attempt to remove the remaining contamination $\delta \lambda_1 \psi_1$ from the first eigenfunction. If we define the new eigenfunction iterates as
\[
\psi_{2new}=\lambda_2 \psi_2 + \delta \lambda_1 \psi_1
\]
\[
\psi_{1new}=\lambda_1 \psi_1 + \epsilon \lambda_2 \psi_2
\]
then
\[
\psi_{2new}(\eta) \leftarrow
  \Bigl(
  \psi_{2new} 
 +\eta_2 \psi_{1new} 
 \Bigr)
  + \eta \psi_{1new}
\]
Thus, the effect of Eq.~(\ref{eq:further_refinement}) is promoting the convergence of $\hat\psi'$ to the first eigenfunction in the normal power method way whereas $\hat\psi''$ (the second eigenfunction estimate) is being corrected at each step by adding (removing) a little of the first eigenfunction estimate. Convergence is reached when $\eta_2 \to 0$; that is, when the second eigenfunction needs no correction from the first eigenfunction.

The above analysis leads to a simple numerical algorithm. The basic steps are
\begin{description}
\item[Step 1:] Initialize
\begin{enumerate}
\item Set convergence parameter $\epsilon$ to a small value.
\item Choose initial estimates $\psi' \approx \psi_1$ and $\psi'' \approx \psi_2$
\end{enumerate}
\item[Step 2:] Reset
\begin{enumerate}
\item Normalize $\psi'\leftarrow \psi' / \|\psi'\|$ and 
$\psi''\leftarrow \psi'' / \|\psi''\|$
\end{enumerate}
\item[Step 3:] Execute power step
\begin{enumerate}
\item Apply $A$ to $\psi'$ and $\psi''$ and solve resulting quadratic balance condition (Eq. (\ref{eq:balance}) )
\item If the roots are real, assign the  roots  $\eta_1$ and $\eta_2$ to correspond to the largest and smallest (in magnitude) eigenvalue estimates respectively and then update via
\begin{eqnarray*}
\psi'  &\leftarrow& A \psi' \\
\psi'' &\leftarrow& A \psi''+\eta_2 A \psi'
\end{eqnarray*}
else update via
\begin{eqnarray*}
\psi' &\leftarrow& A \psi' \\
\hat\psi'' &\leftarrow& A \hat\psi''
\end{eqnarray*}
\end{enumerate}
\item[Step 4:] Test for convergence
\begin{enumerate}
\item If $|\eta_2| > \epsilon$, go to Step 2
\end{enumerate}
\item[Step 5:] Terminate
\end{description}
Eigenvalue estimates can be made by placing $\psi'$ and $\psi''$ in Eq. (\ref{eq:groupings}) for the same or different $R_i$. These $R_i$ can be the same or different from the two used to compute the $q_i$. When the roots are complex, an alternative to Step 3.2 is the use of complex arithmetic.  If it is used, then the $\psi'$ and $\psi''$ estimates in Step 3.2 are updated with complex $\eta_i$.

\subsection{Second Refinement}
Because  Eq. (\ref{eq:nth_step} ) will be almost true for large $n$, it yields a way to estimate the $\psi_i$. If $C_i$ are normalizing constants and  $\eta_2$ is the root that gives $\lambda_2$, then from Eq. (\ref{eq:nth_step}) the $(n+1)^{th}$ guess at $\psi_2$ is
\[
\psi_2 \approx  \psi_2^{(n+1)} = C_2 \Bigr[A^n \psi \Bigl]{_{\eta=\eta_2}}
\]
If $\eta_1 \approx -a_2/b_2$ is the
root that gives $\lambda_1$, then
from Eq. (\ref{eq:nth_step}) the $(n+1)^{th}$ guess at $\psi_1$ is
\[
 \psi_1 \approx \psi_1^{(n+1)} = C_1 \Bigr[A^n \psi \Bigl]{_{\eta=\eta_1}}
\]
These two estimates suggest using 
\[
 \psi^{(n+1)} = \psi_2^{(n+1)} + \eta \psi_1^{(n+1)}
\]
as the next iteration guess. Next, we insert 
\[
 \hat \psi' = \psi_2^{(n+1)} 
\]
\[
\hat  \psi'' =  \psi_1^{(n+1)}
\]
into Eq. (\ref{eq:balance}) and solve it for the two $\eta$ roots. Now we take for new estimates (with the $C_i$ providing normalization)
\begin{equation}
\psi_2 \approx  \psi_2^{(n+2)} = C_2 \Bigr[A  \psi^{(n+1)} \Bigl]{_{\eta=\eta_2}}
\label{eq:t6}
\end{equation}
\begin{equation}
 \psi_1 \approx \psi_1^{(n+2)} = C_1 \Bigr[A \psi^{(n+1)} \Bigl]{_{\eta=\eta_1}}
\label{eq:t7}
\end{equation}
We note that Eq. (\ref{eq:t6}) is the same adjustment to the second eigenfunction estimate as in the first refinement, which uses the best estimate of $\psi_2$. Equation (\ref{eq:t7}) uses the best estimate of $\psi_1$ instead of the power iterated estimate used in first refinement. Empirically, this second refinement simultaneously produces estimates of $\psi_1$ converging as $\lambda_3 / \lambda_1$ and estimates of $\psi_2$ converging as $\lambda_3 / \lambda_2$. In the appendix we demonstrate these rates of convergence for non-degenerate states.


Incorporating this refinement requires only replacing Step 3 of the algorithm for the previous with
\begin{description}
\item[Step 3:] Execute power step
\begin{enumerate}
\item Apply $A$ to $\psi'$ and $\psi''$ and solve resulting quadratic balance condition (Eq. (\ref{eq:balance})) 
\item If the roots are real, assigning the  roots  $\eta_1$ and $\eta_2$ to correspond to the largest and smallest (in magnitude) eigenvalue estimates respectively and then update via
\begin{eqnarray*}
\psi'  &\leftarrow& A \psi'' + \eta_1 A \psi'\\
\psi'' &\leftarrow& A \psi'' + \eta_2 A \psi'
\end{eqnarray*}
else update via
\begin{eqnarray*}
\psi' &\leftarrow& A \psi' \\
\hat\psi'' &\leftarrow& A \hat\psi''
\end{eqnarray*}
\end{enumerate}
\item[Step 4:] Test for convergence
\begin{enumerate}
\item If $|\eta_2| > \epsilon$, go to Step 2
\end{enumerate}
\item[Step 5:] Terminate
\end{description}

\subsection{Practical Algorithm}

In an actual implementation of these algorithms, monitoring convergence by $|\eta_2|<\epsilon$ is not the only choice. The more common way would be monitoring successive estimates of the $\lambda_i$ plus monitoring the residuals $\parallel A\psi'-\lambda_1\psi'\parallel$ and  $\parallel A\psi''-\lambda_2\psi''\parallel$. We also note the following alternative: As $\psi'$ converges to $\psi_1$ and  $\psi''$ converges to $\psi_2$, $q_0$ and $q_2$ converge to zero. In short, multiple criteria exist, leading to cross checks. Some recycle already computed quantities and are consequently quite efficient. Here is an algorithm for the second refinement suitable for implementation:

\begin{description}
\item[Step 1:] Initialize
\begin{enumerate}
\item Set convergence parameters $\epsilon_0$ and $\epsilon_2$ to small values.
\item Initialize iteration index $n=0$,
\item Choose initial estimates for $\psi'$ and $\psi''$,
\item Choose the rules $R_1$ and $R_2$ for grouping of iterated vector components.
\end{enumerate}
\item[Step 2:] Reset
\begin{enumerate}
\item Normalize $\psi'\leftarrow \psi' / \|\psi'\|$ and 
$\psi'' \leftarrow \psi'' / \|\psi''\|$
\end{enumerate}
\item[Step 3:] Execute power step
\begin{enumerate}
\item Apply $A$ to $\psi'$ and $\psi''$,
\item Solve resulting quadratic balance condition (Eq. \ref{eq:balance}),
\item Estimate eigenvalues using either rule (region) $R_1$ or $R_2$,
\item If the roots are real, assign the roots $\eta_1$ and $\eta_2$ to correspond to the largest and smallest (in magnitude) eigenvalue estimates. For example, we will have
\begin{eqnarray*}
\lambda_1  &=& \frac{ \sum\limits_{\alpha\in R_1}\sum\limits_\beta A_{\alpha\beta}\psi_\beta'
              +\eta_1 \sum\limits_{\alpha\in R_1}\sum\limits_\beta A_{\alpha\beta}\psi_\beta'' }
                    { \sum\limits_{\alpha\in R_1}\psi_\alpha''
               +\eta_1\sum\limits_{\alpha\in R_1}\psi_\alpha'' }\\
\lambda_2  &=& \frac{ \sum\limits_{\alpha\in R_1}\sum\limits_\beta A_{\alpha\beta}\psi_\beta'
              +\eta_2 \sum\limits_{\alpha\in R_1}\sum\limits_\beta A_{\alpha\beta}\psi_\beta'' }
                    { \sum\limits_{\alpha\in R_1}\psi_\alpha''
               +\eta_2\sum\limits_{\alpha\in R_1}\psi_\alpha'' }\\
\end{eqnarray*}
then update via
\begin{eqnarray*}
\psi'  &\leftarrow& A \psi'' + \eta_1 A \psi'\\
\psi'' &\leftarrow& A \psi'' + \eta_2 A \psi'
\end{eqnarray*}
else update via
\begin{eqnarray*}
\psi' &\leftarrow& A \psi' \\
\hat\psi'' &\leftarrow& A \hat\psi''
\end{eqnarray*}
\end{enumerate}
\item[Step 4:] Test for convergence
\begin{enumerate}
\item If either $|q_0|> \epsilon_0$ or $|q_2| > \epsilon_2$, increment the iteration index, $n \leftarrow n+1$ and go to Step 2.\end{enumerate}
\item[Step 5:] Terminate.
\end{description}
When the roots are complex, an alternative to Step 3.2 is the use of complex arithmetic.  If it is used, then the eigenvector estimates in Step 3.2 are updated with complex $\eta_i$. The choice of rules is quite flexible. A rule may use one vector component selected randomly, a small number of components selected randomly, all odd or even components, the first or second half of the vector, etc. We note that $q_2$ goes to zero faster than $q_0$, and when $q_2$ becomes very small, then the quadratic equation numerically reduces to $q_1\eta + q_0=0$ which is solved to get $\eta_2$.  Essentially $q_2=0$ means that the dominant eigenpair is known to machine accuracy, so it cannot be improved on further iteration.

\section{Applications}

\subsection{Cyclic Matrix}

To illustrate the effectiveness of the {\it first} refinement, we applied it to the symmetric $N \times N$ matrix
\[
A=\left[ {\begin{array}{*{20}r}
   2 & { - 1} & 0 &  \cdots  & 0 & { - 1}  \\
   { - 1} & 2 & { - 1} &  \ddots  & {} & 0  \\
   0 &  \ddots  &  \ddots  &  \ddots  &  \ddots  &  \vdots   \\
    \vdots  &  \ddots  &  \ddots  &  \ddots  &  \ddots  & 0  \\
   0 & {} &  \ddots  & { - 1} & 2 & { - 1}  \\
   { - 1} & 0 &  \cdots  & 0 & { - 1} & 2  \\
\end{array}} \right]
\]
whose eigenvalues for any $N$ are
\[
\gamma_n = 2 - 2\cos k_n  = 4\sin^2 \frac{{k_n }}{2}
\]
where $k_n  = \frac{{2\pi n}}{N}$ with $n = 0, 1, 2, \ldots , N-1 $. Physically, the matrix represents the $N$ point discretization of the second derivative defined on a circle. We note that for $N$ odd, all but the minimal eigenstate $(n=0)$ are doubly degenerate, while for $N$ even, all but the minimal $(n=0)$ and maximal $(n=N/2)$ ones are doubly degenerate. Accordingly for even $N$, 
$$
0 = \gamma_0 < \gamma_1=\gamma_{N-1} < \cdots < \gamma_{N/2-1}=\gamma_{N/2+1} < \gamma_{N/2}=4
$$

Table 1 reports the results of a deterministic computation of the second smallest eigenpair for a sequence of even $N$. To generate it, all the eigenvalues of the matrix were shifted by subtracting four times the identity matrix and then getting the two largest magnitude eigenvalues of $A-4I$. For the shifted matrix $\lambda_1=\gamma_0=-4$ and $\lambda_2=\gamma_1=-4\cos^2(\pi/N)$. $\lambda_1$ is thus seen as being independent of $N$ and is not reported. The $\lambda_2$'s in Table 1 are 4 plus the power method's computation of second largest magnitude eigenvalue of $A-4I$. The iteration was stopped when the absolute value of the maximum difference between any component of the eigenvector in successive iterations was less than $10^{-10}$. We see remarkable agreement between the values determined by the power method and the exact analytic value is obtained even for largest possible $N$ on our desktop computer. We converged accurately to the second smallest eigenvalue even though it is approaching the smallest one as $N$ is increased and is itself degenerate. By it being degenerate, our eigenvector estimate is a linear combination of two eigenstates that depends on the starting conditions for the iteration. It is not something we can benchmark but it does approximate well $A\psi=\lambda_2\psi$. For $R_1$ and $R_2$, we used the first and second half of the vector components. For N up to about 1000, starting vectors whose components were set randomly worked well. For starting vectors at $N>1000$ we used the eigenvectors found at $N/2$ injected into the higher dimension via $a(2j) \leftarrow 0.75b(j)+0.25b(j+1)$ and $a(2j+1)\leftarrow 0.25b(j)+0.75b(j+1)$. The coefficients were chosen to adjust for the fact that a(2j) is 1 unit from b(j) and 3 units from b(j+1) and  a(2j+1) is 1 unit from b(j+1) and 3 units from b(j).

%
%
%
%
%

\begin{table}
\caption{For the cyclic matrix, the exact and first refinement calculations for the sub-dominant eigenvalue $\lambda_2$, plus their difference. \label{table1}}
\begin{ruledtabular}
\begin{tabular}{rccc}
         N  &    Exact             & PM1        & Difference \\
       100  &  0.0039465433630297  & 0.0039465431649277 &  1.98E-10 \\
       200  &  0.0009868793234571  & 0.0009868792721619 &  5.13E-11 \\
       400  &  0.0002467350504105  & 0.0002467351362063 & -8.58E-11 \\
       800  &  0.0000616847138537  & 0.0000616847980273 & -8.42E-11 \\
      1600  &  0.0000154212379171  & 0.0000154212887717 & -5.09E-11 \\
      3200  &  0.0000038553131951  & 0.0000038553389818 & -2.58E-11 \\
      6400  &  0.0000009638285310  & 0.0000009638405936 & -1.21E-11 \\
     12800  &  0.0000002409571473  & 0.0000002409625672 & -5.42E-12 \\
     25600  &  0.0000000602392878  & 0.0000000602416170 & -2.33E-12 \\
     51200  &  0.0000000150598221  & 0.0000000150608281 & -1.01E-12 \\
    102400  &  0.0000000037649555  & 0.0000000037654755 & -5.20E-13 \\
    204800  &  0.0000000009412389  & 0.0000000009414407 & -2.02E-13 \\
    409600  &  0.0000000002353098  & 0.0000000002353788 & -6.91E-14 \\
    819200  &  0.0000000000588274  & 0.0000000000589155 & -8.82E-14 \\
   1638400  &  0.0000000000147069  & 0.0000000000148614 & -1.55E-13 \\
   3276800  &  0.0000000000036766  & 0.0000000000036855 & -8.88E-15 \\
\end{tabular}
\end{ruledtabular}
\end{table}

\subsection{Two-Dimensional Ising Model}
 
The two-dimensional Ising model is one of the few two-dimensional models of a system of many interacting degrees of freedom that has an exact solution for its thermodynamic properties. This solution, first constructed by Onsager \cite{onsager}, shows that in the thermodynamic limit the model has a phase transition between an magnetically ordered (ferromagnetic) state at low temperatures and a magnetically disordered state (paramagnetic) at high temperatures. Onsager succeeded in calculating many of the properties of the model exactly, including the temperature $T_c$ at which the transition occurs. Key to his calculations was expressing the partition function of the model in terms of its transfer matrix \cite{montroll}, finding the dominant eigenvalue of this matrix, and showing in the thermodynamic limit (letting the area of the model approach infinity) that this eigenvalue implies the onset of long-range ordering among the spin variables of the model.
 
We will consider the model for finite area, that is, an $m\times n$ model defined with periodic boundary conditions in one direction and open boundary conditions in the other. Because of the one open boundary, the transfer matrix will thus be non-symmetric. In the absence of an applied magnetic field, the model's energy is
\[
E\left\{ \mu  \right\} =  - J\sum\limits_{i = 1}^{m-1} {\sum\limits_{j = 1}^n {\mu _{i,j} \mu _{i,j - 1} } }  -
J\sum\limits_{i = 1}^m {\sum\limits_{j = 1}^n {\mu _{i,j} \mu _{i,j + 1} } }
\]
Here, $(i,j)$ are the coordinates of a lattice site. The Ising spin variable $\mu_{i,j}$ on each site has the value of $ \pm 1$, $J>0$, and $\mu _{i,n + 1}  = \mu _{i,1}$. A column configuration of Ising spins will be denoted by
\[
\sigma _j  = \left( {\mu _{1,j} ,\mu _{2,j} , \ldots ,\mu _{m,j} } \right)
\]
and there are $2^m$ possible configurations for each column.
  
The definition of the transfer matrix follows from the expression for the partition function \cite{thompson}
\[
\begin{array}{rcl}
 Z\left( {m,n} \right) & = &\sum\limits_{\left\{ \mu  \right\}} {\exp \left[ { - \beta E\left( {\left\{ \mu  \right\}} \right)} \right]}  \\
  & = &\sum\limits_{\sigma _1 , \ldots ,\sigma _n } {\exp \left[ { - \beta \left( {\sum\limits_{j = 1}^n
  {\left\{ {V_1 \left( {\sigma _j } \right) + V_2 \left( {\sigma _j ,\sigma _{j + 1} } \right)} \right\}} } \right)} \right]}  \\
  & = & \sum\limits_{\sigma _1 , \ldots ,\sigma _n } {L(\sigma _1 ,\sigma _2 )} L(\sigma _2 ,\sigma _3 )
  \cdots L(\sigma _{n - 1} ,\sigma _n )L(\sigma _n ,\sigma _1 ) \\
  & = &\sum\limits_{\sigma _1 } {L^n (\sigma _1 ,\sigma _1 )}  \\
 \end{array}
\]
where
\[
V_1 \left( {\sigma _j } \right) =  - \nu \sum\limits_{i = 1}^{m - 1} {\mu _{i,j} \mu _{i + 1,j} }
\]
is the interaction energy of the $j^{th}$ column and
\[
V_2 \left( {\sigma _j ,\sigma _{j + 1} } \right) =  - \nu \sum\limits_{i = 1}^m {\mu _{i,j} \mu _{i,j + 1} }
\]
is the interaction energy between the $j^{th}$ and $(j+1)^{th}$ columns, $\nu=J/k_B T$, $k_B$ is Boltzmann's constant, and $L(\sigma,\sigma')$ is the transfer matrix of order $N=2^m \times 2^n$ whose elements are
\[
L \left( \sigma ,\sigma'\right) = \exp \left(\nu \sum\limits_{k = 1}^{m -1}
\mu _k \mu _{k + 1} \right)  \exp \left( \nu \sum\limits_{k = 1}^m
\mu _k \mu _k^{'} \right)
\]
More succinctly,
\[
Z(m,n) = {\rm{Tr}}\left( {{\rm{L}}^{\rm{n}} } \right) = \sum\limits_{j = 1}^{2^m } {\lambda _j^n }
\]
 
Onsager found analytic expressions for all the eigenvalues of the transfer matrix. Since we needed to form this matrix, we found it as convenient to compute them numerically. Spot checks produce excellent agreement between the the two approaches. In the thermodynamic limit, when $T\rightarrow T_c$, $\lambda_2\rightarrow \lambda_1$. Here, although we chose $T=T_c$ and the order of or matrix became quite large, we were still reasonably far away for this critical point. Table~\ref{table2} presents a comparison of the two largest eigenvalues of the transfer matrix as determined by our second refinement of the power method and those determined by the EISPACK eigensolver RG \cite{eispack}. For cases except $m=11$, we simply used as the solution the results  after 100 iterations. For $m=11$, we used 1000 iterations. The larger number of iterations was necessary to obtain the same level of  accuracy.

\begin{table}
\caption{Comparison of the two largest eigenvalues of the transfer matrix for the 2D Ising model as a function of the lattice edge size $m$ computed by the EISPACK routine RG and the second refinement. $N=2^m$ is the order of the matrixAlso shown is the fractional difference $(\lambda_{RG}-\lambda_{PM2})/\lambda_{RG}$ between the different estimates. We choose $m=n$ and $T=T_c$.\label{table2}}
\begin{ruledtabular}
\begin{tabular}{rrccc}
m &  N &    RG         &         PM2        &  FD \\
 1 &       4 & 3.41421355573626E+00 & 3.41421355573626E+00 & 0.00E+00 \\
   &         & 1.41421355573626E+00 & 1.41421355573626E+00 & 0.00E+00 \\
 2 &      16 & 7.46410158611908E+00 & 7.46410158611907E+00 & 3.57E-16 \\
   &         & 4.82842709270073E+00 & 4.82842709270073E+00 &-1.29E-15 \\
 3 &     64  & 1.78770541980345E+01 & 1.78770541980345E+01 &-7.95E-16 \\
   &         & 1.35518083939891E+01 & 1.35518083939891E+01 & 3.93E-16 \\
 4 &    256  & 4.41298558292434E+01 & 4.41298558292434E+01 & 4.83E-16 \\
   &         & 3.60398703210879E+01 & 3.60398703210878E+01 & 5.91E-16 \\
 5 &    1024 & 1.10192319565854E+02 & 1.10192319565854E+02 & 9.03E-16 \\
   &         & 9.38962258961220E+01 & 9.38962258961221E+01 &-4.54E-16 \\
 6 &    4096 & 2.76599914093667E+02 & 2.76599914093667E+02 &-8.22E-16 \\
   &         & 2.42266413140723E+02 & 2.42266413140723E+02 & 0.00E+00 \\
 7 &   16384 & 6.96269201662783E+02 & 6.96269201662782E+02 & 1.47E-15 \\
   &         & 6.21748520715910E+02 & 6.21748520715909E+02 & 1.65E-15 \\
 8 &   65536 & 1.75565374661531E+03 & 1.75565374661531E+03 & 1.30E-16 \\
   &         & 1.59043428137424E+03 & 1.59043428136461E+03 & 6.05E-12 \\
 9 &  262144 & 4.43180239838645E+03 & 4.43180239838646E+03 &-1.44E-15 \\
   &         & 4.05958858259757E+03 & 4.05958858259756E+03 & 1.79E-15 \\
10 & 1048576 & 1.11957434253463E+04 & 1.11957434253463E+04 & 1.46E-15 \\
   &         & 1.03466429299731E+04 & 1.03466429299731E+04 & 8.79E-16 \\
11 & 4194304 & 2.82985308867953E+04 & 2.82985308867954E+04 &-3.60E-15 \\
   &         & 2.63419326613631E+04 & 2.63419326613632E+04 &-3.87E-15 \\
\end{tabular}
\end{ruledtabular}
\end{table}

\subsection{One Dimensional Hubbard Model}

The Hubbard Hamiltonian was originally proposed as a model for metallic ferromagnetism \cite{hubbard}. Most recently, its two-dimensional version has been the subject of intense scrutiny as a possible model for electronic superconductivity. In one-dimension, a variant of it, called the Pariser-Parr-Popple Hamiltonian is frequently used to model conjugated cyclic molecules. Other variants model one-dimensional organic conductors. Because of the enormous amount of computer memory required by deterministic methods, precise specification of the ground state (zero temperature) properties of these models has often been hampered by techniques limited to relatively small system sizes. The memory requirements scale as $4^N$, where $ N $ is the number of lattice sites. 
 
The Hamiltonian operator for the Hubbard model is
\[
\hat H =  - t\sum\limits_{\left\langle {i,j} \right\rangle ,\sigma } {\left( {\hat c_{i,\sigma }^\dag  \hat c_{j,\sigma }  + \hat c_{j,\sigma }^\dag  \hat c_{i,\sigma } } \right)}
 + U\sum\limits_i {\hat n_{i, \uparrow } \hat n_{i, \downarrow } }
\]
where the summation is over nearest-neighbor pairs of lattice sites $i$ and $j$ and electron spin $\sigma$; $t$ and $U$ are the hopping amplitude and repulsive Coulomb parameters; ${\hat c_{i,\sigma }^\dag  }$, ${\hat c_{i,\sigma } }$ and $\hat n_{i,\sigma} =\hat c_{i,\sigma }^\dag  \hat c_{i,\sigma}$ are the creation, destruction, and number operators for an electron at site $i$ with spin $\sigma$. Usually a Fock basis is used to represent the Hamiltonian operator as a matrix
\[
h_{ij}  = \left\langle i \right|\hat H\left| j \right\rangle 
\]
where
\[
\left| i \right\rangle  = \left| {n_{1,\uparrow}  ,n_{2,\uparrow}  , \ldots ,n_{N,\uparrow}  } \right\rangle \left| {n_{1,\downarrow}  ,n_{2,\downarrow}  , \ldots ,n_{N,\downarrow}  } \right\rangle 
\]
with $n_i^\sigma = 0,1$ being the eigenvalues of the number operator.  As a representative of an Hermitian operator whose matrix elements are real, the resulting matrix is symmetric. Various symmetries are usually used to block diagonalize the matrix and then obtain the ground state for each block. We will consider only blocks that have a specific value of the z-component of the total electron spin. The size of the Hilbert space and hence the order of the matrix is
\[
N=\frac{N!}{N_\uparrow!(N-N_\downarrow)!}
\]
where $N_\uparrow$ is the number of up spin electrons and being and $N_\downarrow$ is the number of down spin electrons.

For a given lattice site $i$ the maximum number of non-zero values of $h_{ij}$ is $2zN$ where $z$ is the number of nearest neighbors of the chosen lattice. Typically, $z \ll N$; hence, the matrix is very sparse. Here, we will consider the model in one-dimension where $z=2$. In one-dimension the model has an exact solution. Obtaining the ground or first excited state for these solution is not as straightforward as for the two previous test cases. We chose to obtain them numerically and compare the effectiveness of our second refinement of the power method to that of several standard eigenpair methods. Our emphasis is on how well degenerate states are captured.

We will compute the two largest and two smallest eigenvalues of the sparse, potentially hugely dimensioned, matrix $H$ representing the operator $\hat H $ in one dimension with periodic boundary conditions. In this case and if $N_\uparrow=N_\downarrow$ and $N_\uparrow$ is odd, the model satisfies the following version of the Perron-Frobenius Theorem \cite{wilf}: If a matrix is irreducible and all off-diagonal elements are non-positive, the state corresponding to smallest eigenvalue is real and  non-degenerate. We will study the model on a 10 site lattice with $U=4$ and $t=1$. For this lattice size and filling half or less, the theorem applies to cases $(N_\uparrow,N_\downarrow)=$(1,1), (3,3), and (5,5). They are called closed shell cases and the result of our calculations for them are shown in Table~\ref{table3}. Because of various symmetries, features of the smallest states are reflected in those of the largest ones.

\begin{center}
\begin{table}
\caption{For closed shell cases, comparison of the eigenvalues of a 10 site 1D Hubbard model computed by the eigenpair routine DSYEV, the block Lanczos routine DNLASO, and the second refinement. For the first two methods, the three largest and three smallest eigenvalues were computed to measure their consistency, effectiveness, and accuracy. \label{table3}}
\begin{ruledtabular}
\begin{tabular}{rrrccc}
$N_\uparrow$ & $N_\downarrow$  & $N$  & DSYEV & DNLASO &  PM2 \\
1 & 1 & 100   &  0.5657693716217906E+01  &   0.5657693716217901E+01  &   0.5657693716217914E+01 \\ 
  &   &       &  0.5519554669107880E+01  &   0.5519554669107876E+01  &   0.5519554669107137E+01 \\
  &   &       & -0.3862202348191250E+01  &  -0.3862202348191248E+01  &  -0.3862202348191251E+01 \\
  &   &       & -0.3618033988749895E+01  &  -0.3267468797160054E+01  &  -0.3618033988603501E+01 \\
3 & 3 & 14400 &  0.1656339684606611E+02  &   0.1656339684376816E+02  &   0.1656339684606624E+02 \\      
  &   &       &  0.1617312172182284E+02  &   0.1617312172136987E+02  &   0.1617312172191405E+02 \\  
  &   &       & -0.8262531385370846E+01  &  -0.8262531383972004E+01  &  -0.8262531385370927E+01 \\  
  &   &       & -0.7599976793651736E+01  &  -0.7599976793264113E+01  &  -0.7599976793831864E+01 \\
5 & 5 & 63504 &                          &   0.2583432263352126E+02  &   0.2583432263577081E+02 \\
  &   &       &                          &   0.2543485463377173E+02  &   0.2543485464252857E+02 \\
  &   &       &                          &  -0.5834322635176973E+01  &  -0.5834322635773042E+01 \\
  &   &       &                          &  -0.5434854632148166E+01  &  -0.5434830052960784E+01 \\
\end{tabular}
\end{ruledtabular}
\end{table}
\end{center}

In Table~\ref{table3}, three methods where used to get the eigenvalues. One method used was the LAPACK routine DSYEV \cite{lapack}. This double precision routine returns all the eigenvalues and eigenvectors of a symmetric matrix.  At large orders computer memory became insufficient for its use.Accordingly, we supplemented our results with those obtained by using the DNLASO double precision subroutine \cite{netlib} which is a block Lanczos method with selective reorthogonalization \cite{golub}. The components of the starting vectors for the Lanczos iteration are selected randomly and uniformly on the interval (-0.5,0.5). The quality of the results is controlled by specifying the block size (the number of starting states), the number of significant figures for the convergence of the eigenvalues, and the maximum number of iterations. We found a block size of 1 gave estimates for the second and third eigenvalues that became progressively poorer as $N$ increased. This is reasonable. A size of 2 produced cases where the sub-dominant eigenvalue was consistently returned as the dominant. For a size of 6, convergence was very slow if at all. For a size of 8, memory soon became insufficient. A size of 4 was used for the data in the table. We found lack of convergence for many cases if the precision was requested to be larger than 8 decimal places. Typically, a few hundred iterations were needed, but the computation times were a few tens of seconds.  Table~\ref{table3} shows excellent agreement between all three methods. We note that the excited state was at least doubly degenerate.

The next set of results are for electron fillings where the eigenstates are subjected to Kramers degeneracies.  Kramers's Theorem \cite{tinkham} says all energy levels of a system containing an odd number of electrons must at least be at least doubly degenerate provided there are no magnetic fields present to remove time-reversal symmetry. In Table~\ref{table4}, we present several Kramers cases where $(N_\uparrow,N_\downarrow)=$ (3,2), (4,3), and (5,4). For the standard software packages, we listed the three largest and three smallest eigenvalues they estimated to see the accuracy to which they determined the degenerate ground state. We see very good agreement between all three methods in estimating the eigenvalue of degenerate largest and smallest state.  All three however lack the precision necessary to differentiate between a true degeneracy and a very near one. The power method in particular is less than adequate for this purpose.

In Table~\ref{table4}, the three largest and smallest eigenvalues are presented to provide extra information about the degeneracies. Sixteen significant figures were printed to indicate how well degeneracies are captured. Basically we do not know how the precision of the eigenvalues other then it is no more than difference between eigenvalues that should exactly be degenerate. All the expected features of the lowest eigenstates with regard to degeneracies are exhibited. Because the model has particle-hole symmetry similar features also exists for its largest eigenstates. 

\begin{center}
\begin{table}
\caption{For Kramers degeneracies cases, comparison of the eigenvalues of a 10 site 1D Hubbard model computed eigenpair routine DSYEV, the block Lanczos program DNLASO, and the second refinement of the power method. For the first two methods, the three largest and three smallest eigenvalues were computed to measure their consistency, effectiveness, and accuracy in computing degenerate eigenvalues. \label{table4}}
\begin{ruledtabular}
\begin{tabular}{rrrccc}
$N_\uparrow$ & $N_\downarrow$  & $N$  & DSYEV & DNLASO &  PM2 \\
3 & 2 & 5400  &  0.1306499556833340E+02  &   0.1306499556833341E+02  &   0.1306499556833335E+02 \\ 
  &   &       &  0.1306499556833336E+02  &   0.1306499554321960E+02  &   0.1306499556833335E+02 \\
  &   &       &  0.1282579739183819E+02  &   0.1282579739163641E+02  &                          \\
  &   &       & -0.7511951740365890E+01  &  -0.7511951740281242E+01  &  -0.7511951740365513E+01 \\
  &   &       & -0.7511951740365851E+01  &  -0.7511951731564561E+01  &  -0.7511951740365509E+01 \\
  &   &       & -0.7249884543021683E+01  &  -0.7249884543021723E+01  &                          \\
4 & 3 & 25200 &  0.1816344283994604E+02  &   0.1816344283994595E+02  &   0.1816344283994610E+02 \\
  &   &       &  0.1816344283994604E+02  &   0.1816344283994549E+02  &   0.1816344283994610E+02 \\
  &   &       &  0.1771746494384758E+02  &   0.1771746494296794E+02  &                          \\
  &   &       & -0.8030089029893539E+01  &  -0.8030089029893475E+01  &  -0.8030089030622399E+01 \\
  &   &       & -0.8030089029893492E+01  &  -0.8030089029485101E+01  &  -0.8030089030532327E+01 \\
  &   &       & -0.7521441552342070E+01  &  -0.7521441551092671E+01  &                          \\
5 & 4 & 52920 &                          &   0.2285321122055267E+02  &   0.2285321221296537E+02 \\  
  &   &       &                          &   0.2285321121726444E+02  &   0.2285321166226352E+02 \\
  &   &       &                          &   0.2221382316719896E+02  &                          \\
  &   &       &                          &  -0.6853211221744196E+01  &  -0.6853211215825024E+01 \\
  &   &       &                          &  -0.6853211221310334E+01  &  -0.6853211214979680E+01 \\
  &   &       &                          &  -0.6213823170572150E+01  &                          \\
\end{tabular}
\end{ruledtabular}
\end{table}
\end{center}

For other electron fillings {\it a priori} exact information about degeneracies is lacking. What is known is that when $U=0$ the ground state for most fillings is degenerate. These fillings typically are called open shell cases. When $U\neq 0$, the degeneracies are typically lifted, the degree to which it is however depends on the closeness of the nearest unoccupied eigenstate.  For small systems, the lifting might be minor, and for limited precision calculations this can make distinguishing degenerate and nearly degenerate states difficult. For the results in Table~\ref{table5}, this discussion applies to the cases $(N_\uparrow,N_\downarrow)=$ (2,2) and (4,4) illustrated there. The qualitative character of the results are similar to those of the Kramers's cases.

\begin{center}
\begin{table}
\caption{For open shell cases, comparison of the eigenvalues of a 10 site 1D Hubbard model computed by the LAPACK routine DSYEV, the Netlib program DNLASO, and the second refinement of the power method. For the first two methods, the three largest and three smallest eigenvalues were computed to measure their consistency, effectiveness, and accuracy in computing degenerate eigenvalues. \label{table5}}
\begin{ruledtabular}
\begin{tabular}{rrrccc}
$N_\uparrow$ & $N_\downarrow$  & $N$  & DSYEV & DNLASO &  PM2 \\
2 & 2 & 2025  &  0.1121466372028744E+02  &   0.1121466372028743E+02  &   0.1121466372028747E+02 \\ 
  &   &       &  0.1096186919469933E+02  &   0.1096186919469927E+02  &   0.1096186919053599E+02 \\
  &   &       &  0.1096186919469928E+02  &   0.1096186919469929E+02  &                          \\
  &   &       & -0.6601239688910290E+01  &  -0.6431629846631373E+01  &  -0.6601239688910274E+01 \\
  &   &       & -0.6431629846631359E+01  &  -0.6424903541072491E+01  &  -0.6431629865616197E+01 \\
  &   &       & -0.6431629846631350E+01  &  -0.5854101965765123E+01  &                          \\
4 & 4 & 44100 &                          &   0.2143485463460406E+02  &   0.2143485463565059E+02 \\ 
  &   &       &                          &   0.2106806509093548E+02  &   0.2106806509410040E+02 \\  
  &   &       &                          &   0.2106806508923771E+02  &                          \\
  &   &       &                          &  -0.7647179205940428E+01  &  -0.7647179208191599E+01 \\
  &   &       &                          &  -0.7538791441444121E+01  &  -0.7538797509518616E+01 \\
  &   &       &                          &  -0.7538791440796984E+01  &                          \\
\end{tabular}
\end{ruledtabular}
\end{table}
\end{center}

The results of the three tables indicate that this test case is nontrivial. We judge our preliminary results as indicating that the block Lanczos and the second refinement of the power method have comparable effectiveness. The version of the power method used here lacks the ability to return more than two eigenvalues in contrast to the block Lanczos used that should estimate well at least four.

We comment that the Lanczos method is not a black box. To ensure that the result is the minimum as opposed to some excited state, the calculation usually needs to be run multiple times with different random number seeds or some other means to change the starting vectors, and then the results need to be studied to identified those to be regarded as estimates of the ground state \cite{lin,wang}. An error is usually estimated from the variance of the average of the ground state estimates. We restarted the Lanczos calculations for a few of the cases multiple times. As our interest was qualitative, we did not perform an error analysis but instead presented representative results.

The power method is also not a black box. It used the same sparse matrix as the block Lanczos method. From it we only show the two largest and two smallest as our double precision code did not incorporate a procedure to allow the determination of the third eigenpair. The components of one starting state were selected uniformly and randomly over (0,1); the other from (-0.5,0.5). We defined our regions in the following manner. First we performed a random permutation of the vector components. Region 1 was the first $N/2+1$ of these permuted components; region 2, the last $N/2+1$. Initially, we used $\epsilon_0=10^{-13}$ and $\epsilon_2=10^{-10}$. Multiple starts with changing random number seeds were used to ensure consistency. We first computed the largest two eigenvalues and used the largest value, truncated to two significant figures, as the shift to compute the two smallest. Cases where the lowest or the largest had a near degeneracy converged only so far. Often with $|q_0|<\epsilon_0$ being satisfied while the $\epsilon_2$ being too small for $q_2$ to converge. Instead of adjusting these stopping criteria, we found that simply stopping the iteration after some fixed number of iterations and the choosing the result by locating when the residual $\|A\psi''-\lambda_2\psi''\|$ ceased decreasing was very effective.

\section{Concluding Remarks}

We presented two refinements of the power method that enable the simultaneous determination of two extremal eigenpairs of a matrix. We illustrated their effectiveness by benchmarking them on three quite distinct but physically challenging problems. For the cyclic matrix, we exactly knew the eigenvalues and their degeneracies. We showed we could determine the two smallest extremal eigenvalues to nearly machine precision. For the transfer matrix of the two-dimensional Ising model, we knew the exact values. The two-dimensional Hubbard model was more challenging: As a function of the electron filling various degeneracies exist.  Here, we choose to compare the effectiveness of the second refinement with two standard numerical determinations of the ground and first excited state, illustrating the limitations of all three methods especially when the dominant state is degenerate or very nearly so. In general, the second refinement appears as effective as a readily available implementation of the block Lanczos method with selective reorthogonalization. 

All our test cases involved real matrices, but preliminary testing indicates that the techniques presented also work well for complex matrices and non-symmetric matrices with complex eigenvalues. The techniques presented are easily adapted to the determination of just the dominant or just the sub-dominant eigenpair. Convergence to the dominant one is accelerated as it is controlled by $\lambda_3/\lambda_1$ is instead of $\lambda_2/\lambda_1$. Generalizations to more than two extremal pairs are possible. Preliminary testing for up to four have been promising.

One advantage of our refinements is that they maintain the simplicity of the basic power method. Another is their adaptability to Monte Carlo implementations. In a number of fields of physics and chemistry, the power method is the core of the Monte Carlo methods for determining the ground state of models whose complexity grow exponentially with physical size. Here, the ground state energy is estimated from samplings of the ground state wavefunction. Such samplings may involve only a small fraction of all possible components of the state, and a mixed estimator \cite{hammond} for the energy is often used. Here we pushed the use of such estimators a step further with novel consequences. In two other papers, we will describe Monte Carlo implementations of our refinements and their applications to the transfer matrix of the two-dimensional Ising model and to Hamiltonian matrix of the two-dimensional Hubbard model \cite{booth3}.

The work was supported by the U.~S.~Department of Energy under the LANL/LDRD program.


\appendix*
\section{}

We now show that if the eigenstates are non-degenerate, our second refinement of the power method simultaneously produces estimates of $\psi_1$ converging as $\lambda_3 / \lambda_1$ and estimates of $\psi_2$ converging as $\lambda_3 / \lambda_2$. Suppose that the estimates of $\psi_1$ and $\psi_2$ are very good with only a small mixtures of other components; that is, there is some small $v$ such that
\begin{equation}
\psi_1 \approx  ( 1+d v) \psi_1 + e v \psi_2 + f v \psi_3
\label{eq:t57}
\end{equation}
\begin{equation}
\psi_2 \approx a v \psi_1 +(1+ b v) \psi_2 + c v \psi_3
\label{eq:t58}
\end{equation}
Hence
\begin{equation}
\psi=(a v \psi_1 +(1+ b v) \psi_2 + c v \psi_3)  + x (( 1+d v) \psi_1 + e v \psi_2 + f v \psi_3)
\label{eq:it1}
\end{equation}
Define the total $i^{th}$ eigenfunction component in region $j$ as
\begin{equation}
N_{ij}=\sum_{\alpha \in R_j} \psi_{i,\alpha}
\label{eq:p2.1}
\end{equation}
We now apply the balance condition for equal $\lambda$'s. 
For region $R_1$ the eigenvalue estimate is
$$
\frac{N_{11} \lambda_1 ( a v + x ( 1+d v)) + N_{21} \lambda_2 ((1+ b v) + x e v)+ N_{31} \lambda_3 (c v + x f v)}
{N_{11} ( a v + x ( 1+d v)) + N_{21} ((1+ b v) + x e v)+ N_{31} (c v + x f v)}
$$
while for region $R_2$ it is
$$
\frac{N_{12} \lambda_1 ( a v + x ( 1+d v)) + N_{22} \lambda_2 ((1+ b v) + x e v)+ N_{32} \lambda_3 (c v + x f v)}
{N_{12} ( a v + x ( 1+d v)) + N_{22} ((1+ b v) + x e v)+ N_{32} (c v + x f v)}
$$
Set the above two eigenvalue estimates equal and cross multiply to clear the denominators
 and obtain a quadratic equation in $x$.
We now collect powers of $x$. Terms involving $x^0$ are:
\begin{eqnarray}
L_1 &=&-a \lambda_1 N_{12} N_{21} v + a \lambda_2 N_{12} N_{21} v + a \lambda_1 N_{11} N_{22} v - a \lambda_2 N_{11} N_{22} v \\ \nonumber
L_2&=&-c \lambda_2 N_{22} N_{31} v + c \lambda_3 N_{22} N_{31} v + c \lambda_2 N_{21} N_{32} v - c \lambda_3 N_{21} N_{32} v \\ \nonumber
L_3&=&v^2 (-a b \lambda_1 N_{12} N_{21}+ a b \lambda_2 N_{12} N_{21}+a b \lambda_1 N_{11} N_{22}) \\ \nonumber
L_4&=&-c \lambda_2 N_{22} N_{31} v + c \lambda_3 N_{22} N_{31} v + c \lambda_2 N_{21} N_{32} v - c \lambda_3 N_{21} N_{32} v \\ \nonumber
L_5&=&v^2(-b c \lambda_2 N_{22} N_{31}+b c \lambda_3 N_{22} N_{31}+a c \lambda_1 N_{11} N_{32}) \\ \nonumber
L_6&=&v^2(-a c \lambda_3 N_{11} N_{32}+b c \lambda_2 N_{21} N_{32}-b c \lambda_3 N_{21} N_{32}) 
\label{eq:l1_l6}
\end{eqnarray}
Terms involving $x^1$ are
\begin{eqnarray}
L_7&=&-f \lambda_2 N_{22} N_{31} v + f \lambda_3 N_{22} N_{31} v + f \lambda_2 N_{21} N_{32} v-f \lambda_3 N_{21} N_{32} v \\ \nonumber
L_8&=&v^2(-a e \lambda_1 N_{12} N_{21}+a e \lambda_2 N_{12} N_{21}) \\ \nonumber
L_9&=&v^2(a e \lambda_1 N_{11} N_{22}-a e \lambda_2 N_{11} N_{22}-a f \lambda_1 N_{12} N_{31}) \\ \nonumber
L_{10}&=&v^2(a f \lambda_3 N_{12} N_{31}-c e \lambda_2 N_{22} N_{31}-b f \lambda_2 N_{22} N_{31}) \\ \nonumber
L_{11}&=&v^2(c e \lambda_3 N_{22} N_{31}+b f \lambda_3 N_{22} N_{31}+a f \lambda_1 N_{11} N_{32}) \\ \nonumber
L_{12}&=&v^2(-a f \lambda_3 N_{11} N_{32}+c e \lambda_2 N_{21} N_{32}+b f \lambda_2 N_{21} N_{32}) \\ \nonumber
L_{13}&=&v^2(-c e \lambda_3 N_{21} N_{32}-b f \lambda_3 N_{21} N_{32}) \\ \nonumber
L_{14}&=&(1+ d v)(-\lambda_1 N_{12} N_{21}+\lambda_2 N_{12} N_{21}  + \lambda_1 N_{11} N_{22}  - \lambda_2 N_{11} N_{22} ) \\ \nonumber
L_{15}&=&v(1+ d v)(-b \lambda_1 N_{12} N_{21}   + b \lambda_2 N_{12} N_{21} +b \lambda_1 N_{11} N_{22} -b \lambda_2 N_{11} N_{22} ) \\ \nonumber
L_{16}&=&v(1+ d v)(-c \lambda_1 N_{12} N_{31} +c \lambda_3 N_{12} N_{31}+c \lambda_1 N_{11} N_{32}-c \lambda_3 N_{11} N_{32})
\label{eq:l7_l16}
\end{eqnarray}
Terms involving $x^2$ are
\begin{eqnarray}
L_{17}&=&v^2 e f(- \lambda_2 N_{22} N_{31}+\lambda_3 N_{22} N_{31}+ \lambda_2 N_{21} N_{32}- \lambda_3 N_{21} N_{32}) \\ \nonumber
L_{18}&=&e  v (1+d v)(- \lambda_1 N_{12} N_{21}+ \lambda_2 N_{12} N_{21}+\lambda_1 N_{11} N_{22}-\lambda_2 N_{11} N_{22}) \\ \nonumber
L_{19}&=&f  v (1+d v)(-\lambda_1 N_{12} N_{31}+\lambda_3 N_{12} N_{31}+\lambda_1 N_{11} N_{32}-\lambda_3 N_{11} N_{32})
\label{eq:l17_l19}
\end{eqnarray}
Finally we can write
\begin{eqnarray}
q_0&=&L_1 + L_2 + L_3 + L_4 + L_5 + L_6 \\ \label{eq:q0}
q_1&=&L_7 + L_8 + L_9 + L_{10} + L_{11} + L_{12} +L_{13}+L_{14}+L_{15}+L_{16} \\ \label{eq:q1}
q_2&=&L_{17}+L_{18}+L_{19} \label{eq:q2}
\end{eqnarray}
and note that we seek the smallest magnitude root of
\begin{equation}
q_2 x^2+ q_1 x+ q_0=0
\label{eq:quadx}
\end{equation}
Note that $x$ will be very small when $v\approx 0$, so that the quadratic part can be ignored, leading to
\begin{equation}
x=-q_0/q_1
\label{eq:t65}
\end{equation}
Also the terms involving $v^2$ can be ignored compared to terms involving $v$. Thus
\begin{eqnarray*}
q_0&=&L_1+L_2 \\
 &=&v(-a \lambda_1 N_{12} N_{21}  + a \lambda_2 N_{12} N_{21}  + a \lambda_1 N_{11} N_{22}  - a \lambda_2 N_{11} N_{22} \\
 & &-c \lambda_2 N_{22} N_{31}  + c \lambda_3 N_{22} N_{31}  + c \lambda_2 N_{21} N_{32}  - c \lambda_3 N_{21} N_{32} )
\end{eqnarray*}
Now consider $q_1$. Every term except $L_{14}$ has at least $v^1$ in it, which will be small compared to the $v^0$ term in $L_{14}$. Thus using the $v^0$ term in $L_{14}$,
$$
q_1=(-\lambda_1 N_{12} N_{21}+\lambda_2 N_{12} N_{21}  + \lambda_1 N_{11} N_{22}  - \lambda_2 N_{11} N_{22} )
$$
so from Eq.~(\ref{eq:t65}) we have that
$$
x=-a v - c v \frac{(- \lambda_2 N_{22} N_{31}  +  \lambda_3 N_{22} N_{31}  +  \lambda_2 N_{21} N_{32}  -  \lambda_3 N_{21} N_{32} )}
{(-\lambda_1 N_{12} N_{21}+\lambda_2 N_{12} N_{21}  + \lambda_1 N_{11} N_{22}  - \lambda_2 N_{11} N_{22} )}
$$
In what follows, $\psi$ is replaced by $\psi_2$ because the root corresponding to $\psi_2$ has been selected.
$$
A \psi_2=\lambda_1( a v + x ( 1+d v) )\psi_1+\lambda_2((1+ b v)+x e v) \psi_2 +\lambda_3(c v +x f v)\psi_3
$$
Both $x$ and $v$ are small, so the product  $x v$ is ignored yielding
$$
A \psi_2=\lambda_1( a v + x  )\psi_1+\lambda_2((1+ b v)+x e v) \psi_2 +\lambda_3(c v +x f v)\psi_3
$$
Substituting for $x$, we rewrite this equation as 
$$
A \psi_2=\lambda_1 \frac{- c v(- \lambda_2 N_{22} N_{31}  +  \lambda_3 N_{22} N_{31}  +  \lambda_2 N_{21} N_{32}  -  \lambda_3 N_{21} N_{32} )}
{(-\lambda_1 N_{12} N_{21}+\lambda_2 N_{12} N_{21}  + \lambda_1 N_{11} N_{22}  - \lambda_2 N_{11} N_{22} )}
\psi_1
$$
$$
+\lambda_2(1+ b v) \psi_2 +\lambda_3(c v )\psi_3
$$
Dividing by $\lambda_2(1+ b v)$ and keeping terms to order $v$ yields
$$
\frac{1}{\lambda_2}A \psi_2= \frac{- c v(-  N_{22} N_{31}  + 
( \lambda_3/\lambda_2) N_{22} N_{31}  +   N_{21} N_{32}  - 
 (\lambda_3/\lambda_2) N_{21} N_{32} )}
{(- N_{12} N_{21}+(\lambda_2/\lambda_1) N_{12} N_{21}  + 
 N_{11} N_{22}  -( \lambda_2/\lambda_1) N_{11} N_{22} )}
\psi_1
$$
$$
+ \psi_2 +\frac{\lambda_3}{\lambda_2}(c v )\psi_3
$$
We note that the $\psi_3$ component has dropped by the ratio $\lambda_3/\lambda_2$ after the iteration. Note that the $\psi_1$ component in the above is proportional to  the $\psi_3$ component ($c v$) in the estimate of $\psi_2$ at the beginning of the iteration.  Therefore, both the $\psi_1$ and $\psi_3$ components drop out of the $\psi_2$ estimate as $\lambda_3/\lambda_2$.
 
Now we will look at the estimated $\psi_1$ component. We note that dividing the iterate by a constant does not affect the eigenvalue estimates so we divide (\ref{eq:it1}) by $x$ and then label the estimate as $\psi_1$ because the root corresponding to $\psi_1$ will be chosen.
$$
\psi_1=(a v \psi_1 +(1+ b v) \psi_2 + c v \psi_3) \frac{1}{x} +  (( 1+d v) \psi_1 + e v \psi_2 + f v \psi_3)
$$
The eigenvalue balance equation is the same, but instead of Eq. \ref{eq:quadx}
we have
$$
q_2 +q_1 \frac{1}{x}+ q_0\frac{1}{x^2}=0
$$
When estimating $\psi_1$, note that $1/x$ will be very small when $v\approx 0$, so that the inverse quadratic part of this equation can be ignored, leading to
\begin{equation}
\frac{1}{x}=-q_2/q_1
\label{eq:tlin1}
\end{equation}
After all $v^2$ terms in $q_2$ are dropped, (\ref{eq:q2}) and  (\ref{eq:l17_l19}) become
$$
q_2=e v (- \lambda_1 N_{12} N_{21}+ \lambda_2 N_{12} N_{21}+\lambda_1 N_{11} N_{22}-\lambda_2 N_{11} N_{22})+
$$
$$
  f v (-\lambda_1 N_{12} N_{31}+\lambda_3 N_{12} N_{31}+\lambda_1 N_{11} N_{32}-\lambda_3 N_{11} N_{32})
$$
and after all $v$ and $v^2$ terms in $q_1$ are dropped, (\ref{eq:q1}) and  (\ref{eq:l7_l16}) become
$$
q_1=(-\lambda_1 N_{12} N_{21}+\lambda_2 N_{12} N_{21}  + \lambda_1 N_{11} N_{22}  - \lambda_2 N_{11} N_{22} )
$$
From (\ref{eq:tlin1})
$$
\frac{1}{x}=-e v - \frac{f v (-\lambda_1 N_{12} N_{31}+\lambda_3 N_{12} N_{31}+\lambda_1 N_{11} N_{32}-\lambda_3 N_{11} N_{32})}
{(-\lambda_1 N_{12} N_{21}+\lambda_2 N_{12} N_{21}  + \lambda_1 N_{11} N_{22}  - \lambda_2 N_{11} N_{22} )}
$$
We now note that 
$$
A \psi_1=(a v \lambda_1 \psi_1 +(1+ b v) \lambda_2 \psi_2 + c v \lambda_3 \psi_3) \frac{1}{x} +  (( 1+d v) \lambda_1 \psi_1 + e v \lambda_2 \psi_2 + f v \lambda_3 \psi_3)
$$
If the small terms associated with $v^2$ and $v \frac{1}{x}$ are dropped, 
then dividing the equation by $\lambda_1$ yields (to first order in $v$)
$$
\frac{1}{\lambda_1}A \psi_1= \frac{\lambda_2}{\lambda_1} \psi_2  \frac{1}{x} +  (  \psi_1 + e v
\frac{\lambda_2}{\lambda_1} \psi_2 + f v \frac{\lambda_3}{\lambda_1} \psi_3)
$$
Substituting for $ \frac{1}{x}$,
$$\frac{1}{\lambda_1}A T_1=\psi_1+ f v \frac{\lambda_3}{\lambda_1} \psi_3 +$$
$$
 -f v  \frac{(\lambda_2/\lambda_1)
  (-\lambda_1 N_{12} N_{31}+\lambda_3 N_{12} N_{31}+\lambda_1 N_{11} N_{32}-\lambda_3 N_{11} N_{32})}
{(-\lambda_1 N_{12} N_{21}+\lambda_2 N_{12} N_{21}  + \lambda_1 N_{11} N_{22}  - \lambda_2 N_{11} N_{22} )} \psi_2
$$
We see that the $\psi_3$ component is decreasing as $\lambda_3/\lambda_1$ and that the $\psi_2$ component is proportional to the $\psi_3$ component ($f v$) at the beginning of the iteration. Thus the $\psi_2$ component also should be falling as $\lambda_3/\lambda_1$. Thus the procedure is
converging to the first eigenfunction at the accelerated rate $\lambda_3/\lambda_1$.


\end{document}